**Phase Diffusion in Single-Walled Carbon Nanotube Josephson Transistors**

Y. Zhang[§], G. Liu, C. N. Lau[*]

Department of Physics & Astronomy, University of California, Riverside, CA 92521

[§]Present address: School of Phys. Science & Technology, Southwest University, Chongqing 400715, PR China

We investigate electronic transport in Josephson junctions formed by single-walled carbon nanotubes coupled to superconducting electrodes. We observe enhanced zero-bias conductance (up to $10e^2/h$) and pronounced sub-harmonic gap structures in differential conductance, which arise from the multiple Andreev reflections at superconductor/nanotube interfaces. The voltage-current characteristics of these junctions display abrupt switching from the supercurrent branch to resistive branch, with a gate-tunable switching current ranging from 65 pA to 2.5 nA. The finite resistance observed on the supercurrent branch and the magnitude of the switching current are in good agreement with calculation based on the model of classical phase diffusion.

---

[*] To whom correspondence should be addressed. E-mail address: lau@physics.ucr.edu.

Carbon nanotubes (SWNTs) have emerged as a new model system for quantum dots, as they enjoy several advantages compared with traditional ones based on two-dimensional electron gas (2DEG). For instance, a short nanotube device has relatively large single particle level spacing $\Delta E$ compared to charging energy $E_c$, a parameter regime difficult to attain by traditional methods. Nanotubes can be readily coupled to different electrode materials such as superconductors[1-8] and ferromagnets[9,10], thus enabling investigation of transport of spin and Cooper pairs through 1D nanostructure. A recent report demonstrated that a superconductor/SWNT/ superconductor (S/SWNT/S) Josephson junction (JJ) can function as a gate tunable supercurrent transistor, where supercurrent up to 3 nA was observed[4]. Similar supercurrent branch was also reported in single-walled (SWNT) and multi-walled (MWNT) devices[5,6]. These experimental results inspired much interest in the application of SWNT-based Josephson junctions as a potential building block for quantum computing architectures based on superconductors. Yet, different regimes of the junction dynamics, e.g. underdamped *vs* overdamped junctions, weak *vs* strong Josephson coupling, and how SWNT JJ may differ from conventional junctions in these aspects, have not been explored.

Here we report the observation of gate tunable switching current in single-walled carbon nanotubes coupled to superconducting electrodes. These devices display pronounced peaks in differential conductance at sub-harmonic multiples of $\Delta$, the energy gap of the superconductors. The voltage-current (*V-I*) characteristics exhibit a sharp switching from supercurrent branch to the normal state, with the switching current controllable by gate voltage, changing from 65 pA to 2.5 nA. Both the switching current and the finite resistance observed in the supercurrent branch are consistent with the theoretical prediction using the model of classical phase diffusion[5, 11, 12], in which damping at high frequencies results in the diffusion of the phase difference across the junction down the tilted washboard potential within the framework of resistively and capacitively shunted junctions (RCSJ).

The SWNTs are prepared by chemical vapor deposition method on highly doped $Si/SiO_2$ substrates[13], and located with respect to the predefined alignment marks using an atomic force microscope. The palladium(Pd)/aluminum(Al) electrodes are fabricated by standard electron beam lithography. Only devices with room temperature resistance below 30kΩ are selected for the measurement in a $^3$He refrigerator. This paper presents results from two different devices: Device1 with Pd(6nm)/Al(70nm) bilayer, and source-drain separation of 390nm; Device2 with Pd(3nm)/Al(70nm) bilayer, and separation of 580nm. The gate dependence of room temperature resistance indicates that both SWNTs are both small band gap semiconductors.

Before investigating the superconducting behavior of our S/SWNT/S junctions, the samples were first characterized in normal state by applying a magnetic field $H$=8T that suppresses superconductivity in the Al electrodes. Fig. 1a plots the differential conductance (color) of Device1 as a function of drain-source voltage $V$ (vertical axis) and gate voltage $V_g$ (horizontal axis). The distinct "checker board" pattern, i.e. the sinusoidal oscillation of the device's differential conductance with both gate and bias voltages, arises from the so-called Fabry-Perot (FP) interference[14] of incident and multiply reflected electron waves between two partially transmitting electrodes, or equivalently, from resonant and off-resonant transmission across quantized single particle levels. From Fig. 1a, we can infer that the SWNT/electrode contact is highly transparent, since the device conductance ranges between 1.8 to $3.4e^2/h$, approaching the theoretical limit of $G_0=4e^2/h = (6.5k\Omega)^{-1}$ for a perfectly contacted SWNT. Moreover, the characteristic voltage scale, indicated by the arrow in Fig. 1a, is $V_c\approx4.4$mV. The energy scale $eV_c=hv_F/2$ corresponds to a $2\pi$ modulation in the phase accumulated by an electron in completing a roundtrip between two scatterers separated by distance $L$. Here $v_F\approx8\times10^5$m/s is the Fermi velocity of charges in nanotubes. The value obtained from this measurement $L\approx400$nm is consistent with the source-drain spacing, indicating that scatterings primarily occur at the nanotube-electrode interface, not by defects. Thus, our SWNT devices are relatively free of defects and have almost ohmic contact. For Device 2, similar interference pattern was also observed with an average conductance around $2e^2/h$, albeit not as periodic as Device1, suggesting that Device2 is slightly more disordered.

We now focus on the device behavior with superconducting electrodes at $H=0$. At $|V|\geq50\mu V$, transport was dominated by the quasiparticles and FP interference pattern persists. At small biases $V\leq\pm50\mu V$, the transport characteristics in both devices changed dramatically: conductance peaks are observed, persisting through all gate voltage ranges, indicating enhanced transport through resonant and off-resonant states (Fig. 1b). Note that the conductance around zero bias reaches as high as $10e^2/h \gg G_0$, indicating the superconducting proximity effect. For Device 2 with thinner Pd contact layer, we are able to resolve several pronounced conductance peaks around zero bias. We identify $2\Delta/e=\pm0.15$mV, where the conductance peaks there correspond to the onset of direct quasiparticle transport. The peaks at $\sim V\leq\pm0.075$mV results from multiple Andreev reflection processes[15]. During an Andreev reflection, an incident electron at the SWNT/S interface is reflected as a hole, with the formation of a Cooper pair in superconducting condensate. For a S/SWNT/S junction, an electron can be reflected back and forth between the electrodes several times, each time gaining energy $eV$, before it gathers enough energy to exit SWNT. Such multiple Andreev reflection processes give rise to features in $dI/dV$ at voltages which are sub-harmonic multiples of $2\Delta$[16], and contribute to the giant conductance peak at zero-bias. The above feature of MARs persists throughout the whole range of measured gate voltage, with peaks' position fluctuating slightly with changing $V_g$. In principle, one may expect an infinitely high conductance peak around zero bias, which is an important signature of the Josephson supercurrent, though the actual conductance value measured may be limited due to inelastic scattering inside the SWNT[17].

To investigate the possibility of supercurrent, we current-bias the devices and the dc $V-I$ characteristics are shown in Fig. 3a. At small current $I<\sim$ nA, the devices remains on the supercurrent branch and displays finite (and typically small) resistance; after the bias current exceeds a threshold, $I_s$, the measured voltage abruptly switches to the quasiparticle branch, with a resistance that approaches $R_N$, the normal state resistance at higher currents. Both the switching current $I_s$ and linear branch resistance $R_0$ depends strongly on the gate voltage, and are correlated with the normal state conductance $G_N$ (Fig. 3a inset, 3b): for the gate voltage range between 0.44V to 0.536V, $G_N$ changes from 0.8 to 2.68 $e^2/h$, $I_s$ decreases from about 2.5nA to 65pA, while $R_0$ increases from about 1.5kΩ up to 44kΩ correspondingly. Moreover, there exists a simple relationship between $I_s$ and $R_0$. Fig. 4 inset plots $I_s$ and $R_0$ in logarithmic scales, and the data points fall on a straight line, indicating a power-law dependence. The solid line is a best-fit curve to $I_s = A/R_0$, with the coefficient $A$ given by ~3200 nA·Ω, in good agreement with the data.

Gate-tunable *V-I* characteristics and supercurrent have been observed in SWNT and MWNT[4, 5], which originates from resonant and off-resonant transport across quantized single particle level spacing in a finite SWNT segment. Theoretically, for two superconductors symmetrically coupled via a single discrete energy level with two spin states, the maximum critical current in the wide resonance regime ($\Gamma>>\Delta$) is

$$I_{c0} \approx \frac{2e\Delta}{\hbar}\tanh\left(\frac{\Delta}{2k_BT}\right), \qquad (1)$$

where $T$=300mK is the temperature, $\Gamma$ is the level broadening due to electron's finite lifetime, $e$ is the electron charge, and $k_B$ is Boltzman's constant.[18]. For Device 2, $2\Delta \sim 0.15$meV as determined from MAR features, yielding $I_{c0}$~34.6nA. In realistic devices, the asymmetry in coupling is expected to decrease the measured normal state conductance $G_N$, which in turn leads to reduction in the actual critical current, given by

$$I_c = I_{c0}\left[1 - \sqrt{1 - \frac{G_N}{4e^2/h}}\right] \qquad (2)$$

The maximum $G_N$ is 2.68 $e^2$/h for Device 2, thus we expect the critical current to be as large as ~15 nA. This value is an order of magnitude larger than the observed value of 2.5 nA.

To understand the $I_s \propto R_0^{-1}$ behavior as well as the large discrepancy between theoretical and experimental values of critical current, we focus on the dynamics of the SWNT Josephson junction. We note that *V-I* characteristics of the devices are consistent with that of an underdamped junction[11]. This agrees with a simple estimate of the quality factor within the RCSJ model: $Q=\omega_p R_j C_j$, where $\omega_p = \sqrt{\frac{2eI_c}{\hbar C_j}}$ is the plasma frequency, $R_j$ and $C_j$ are the shunt resistance and capacitance of the junction, respectively. Assuming $R_j$ is given by the quasiparticle resistance $R_N e^{\Delta/kT}$ ~150k$\Omega$ ($R_N$~10k$\Omega$, $\Delta/e$=0.075mV, $T$=0.3K), $C_j$~0.5fF, we obtain $Q\sim30$, indicating that the junction is underdamped with relatively small dissipation. In such junctions, thermal fluctuation tends to prematurely switch the *V-I* characteristics to the resistive branch – the *upper limit* of Josephson energy of the junction is estimated to be $E_J = \frac{\hbar I_c}{2e} \sim$ 0.03meV for $I_c \sim$ 15nA, comparable to the thermal energy $k_BT \sim$ 0.025meV at 300mK, hence thermal activation is expected greatly reduce the measured value of $I_c$. On the other hand, the junction is not isolated from its electromagnetic environment, *e.g.* by inserting high impedance resistors in the leads immediately before the junction. Thus, at the characteristic plasma frequency of the junction, we expect the impedance of the electrical leads to be much smaller than $R_N e^{\Delta/kT}$. Hence, even though the junction is underdamped at low frequencies, it is likely to be overdamped at high frequencies, giving rise to finite zero-bias resistance. We thus seek to quantitatively describe the observed behavior using the model of classical phase diffusion.

In the phase diffusion regime, below the critical current, a finite voltage is measured in the nominally zero resistance state, yielding a zero-bias resistance $R_0$[12]

$$R_0 = \frac{Z}{I_0^2(E_J/k_BT)-1} \qquad (3)$$

where $Z$ is the environmental impedance at dc, typically ~ *50 – 400* Ω, and $I_0(x)$ is the modified Bessel function. Since the upper bound of $E_J$ is comparable with $k_BT$, and the voltage $V$ across the junction is typically only a few μV, we consider the weak Josephson coupling regime of $E_J$, $eV \ll k_BT$, where Equ. (3) is simplified to

$$R_0 = 2Z\left(\frac{k_BT}{E_J}\right)^2 \qquad (4)$$

and the switching current is given by $I_s = \frac{eE_J^2}{2k_BT\hbar}$. Combining the two equations yields

$I_s = \frac{eZk_BT}{\hbar}\frac{1}{R_0}$. Thus, $I_s$ is proportional to $R_0^{-1}$, in agreement with our experimental observation (Fig. 4 inset). Substituting the fitting coefficient $A=3200$ nA·Ω into $\frac{eZk_BT}{\hbar}$, we estimate $Z \sim 485$ Ω, which is a reasonable value. (Here five data points with $R_0 > 1/G_N$ were excluded from the fitting, because the large $R_0$ originates from the opening of gap in the density of states, where Equ. (3) and (4) are applicable.)

Further insight is provided by investigating the dependence of $R_0$ on $G_N$ at different gate voltages. For each $V_g$, we first calculate the ratio $\frac{E_J}{k_BT}$ by solving Equ. (3) numerically using the measured values of $R_0$ and $Z=485$ Ω. These calculated values of $\frac{E_J}{k_BT}$ is then plotted against $G_N$ (in units of $e^2/h$) in Fig. 4. From Equ. (2), we expect

$$\frac{E_J}{k_BT} = \frac{\hbar I_{c0}}{2ek_BT}\left[1 - \sqrt{1 - \frac{G_N}{4e^2/h}}\right] \quad (5)$$

We fit Equ. (5) to the data points to with $I_{c0}$ as the fitting parameter, and obtain reasonable agreement between data (red squares) and calculation (blue line). The value of $I_{c0}$ obtained from fitting is 21 nA, ~ 60% of the ideal value of ~34.6 nA. This reduction is not surprising, considering possible defects in the SWNT segment in Device 2. Thus, our data are well described by the model of phase diffusion in the weak Josephson coupling limit.

In summary, we observe the proximity effect induced superconductivity in S/SWNT/S Josephson junctions, in which the MARs and the supercurrent features can be tuned by the gate voltage. The finite zero bias resistance $R_0$ and magnitude of the switching current $I_s$ in the V-I characteristics are in good agreement with the phase diffusion model in RCSJ.

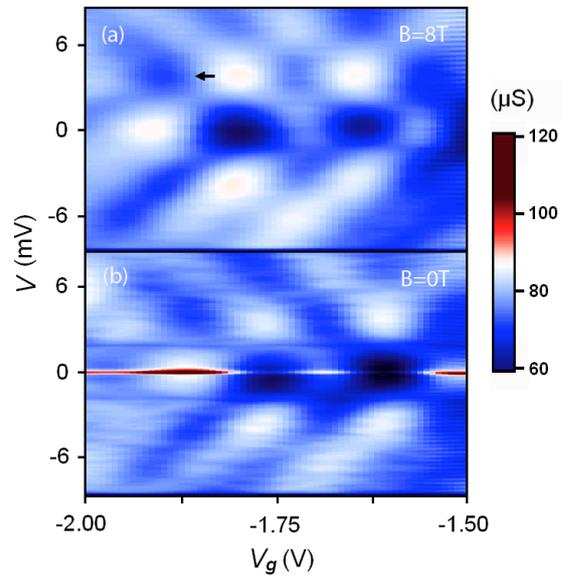

Fig. 1: (color online) Device1: Differential conductance plot as a function of bias and gate voltage (a) with a magnetic field of 8T (b) without magnetic field. The arrow in (a) labeled the characteristic voltage of Fabry-Perot interference.

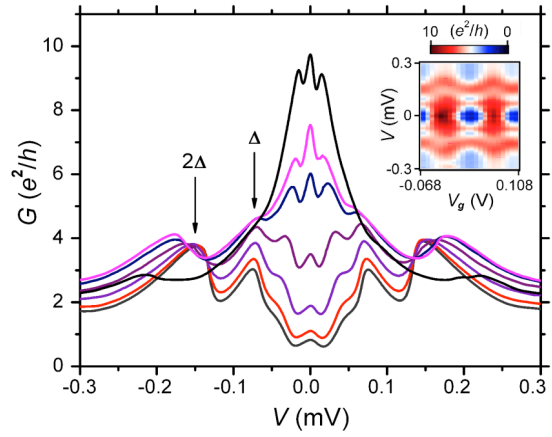

Fig. 2: (color online) Conductance *G* vs. bias voltage *V* at $V_g$= 0.076 - 0.028 in steps of 8mV from top curve to bottom (data from Device 2). The inset shows the MAR plot as a function of bias and gate voltage for two resonant states.

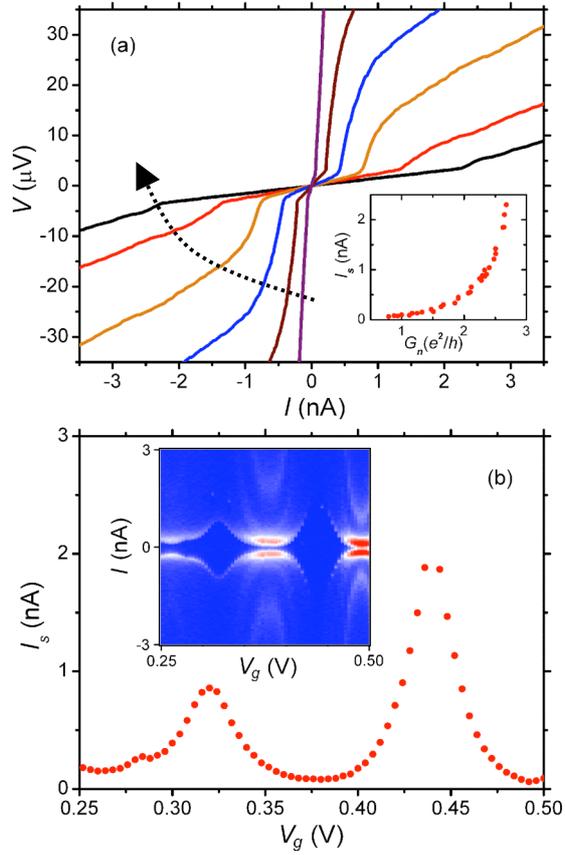

Fig. 3: (color online) (a) *V-I* characteristics show the modulation of the switching current $I_s$ with $V_g$ (in the direction of the arrow, $V_g$ = 0.485, 0.479, 0.473, 0.467, 0.440 V, respectively). The inset displays the switching current vs. normal state conductance $G_N$. (b) Switching current vs. gate voltage at two resonant states. The differential resistance plot was shown as inset.

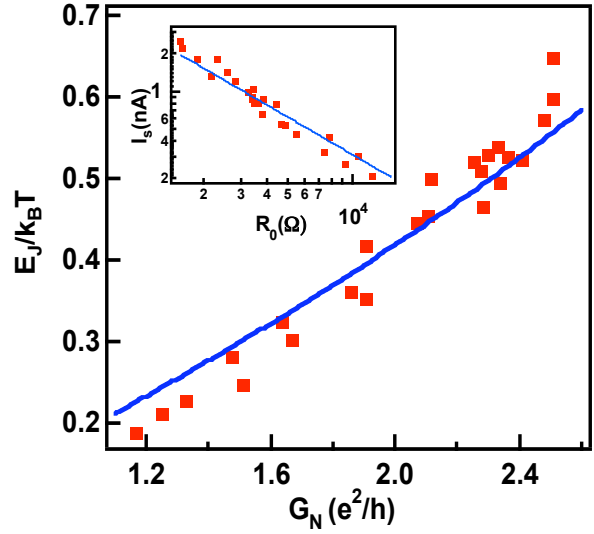

Fig. 4: (color online) Inset: Switching current *vs* zero bias resistance. Red squares: data. The solid line is fit to the data using $I_s = \dfrac{A}{R_0}$, where *A* is a fitting parameter, determined to be 3200 nA·Ω. Main Panel: $\dfrac{E_J}{k_B T}$ (see text) *vs* normal state conductance. The solid line is a fit to Equ. (5).